\documentclass[superscriptaddress,twocolumn,showpacs,preprintnumbers,prb]{revtex4-1}

\usepackage{times,xspace}
\usepackage{amsbsy,amssymb,amsmath,bm}
\usepackage{graphicx,color,epsfig}
\usepackage{fancyhdr}
\usepackage[usenames,dvipsnames,svgnames,table]{xcolor}
\usepackage{hyperref}
\usepackage{ulem}

\usepackage[latin1]{inputenc}
\usepackage{subfig}
\usepackage{mhchem}
\usepackage{hyperref}

\begin{document}

\title{Large magnetic anisotropy in \ce{Fe_{0.25}TaS_2} }
\date{\today}
\author{Vaideesh Loganathan}
\affiliation{Department of Physics and Astronomy, Rice University, Houston, Texas 77005, USA}
\author{Jian-Xin Zhu}
\affiliation{Theoretical Division, Los Alamos National Laboratory, Los Alamos, New Mexico 87545, USA}
\affiliation{Center for Integrated Nanotechnologies, Los Alamos National Laboratory, Los Alamos, New Mexico 87545, USA
}
\author{Andriy H. Nevidomskyy}
\affiliation{Department of Physics and Astronomy, Rice University, Houston, Texas 77005, USA}

\begin{abstract}
	We present a first-principles study of the large magneto-crystalline anisotropy in the intercalated di-chalcogenide material \ce{Fe_{0.25}TaS_2}, investigated with the DFT+U approach. We verify a uniaxial magnetocrystalline anisotropy energy(MAE) of 15meV/Fe. in the material. We further analyze the dependence of MAE on the constituent elements and the effect of spin-orbit coupling. Contrary to conventional intuition, we find a small contribution to MAE due to strong spin-orbit coupling in the heavier element, Ta. We show that the electronic configuration, crystal field environment and correlational effects of the magnetic ion are more important.
\end{abstract}

\maketitle

\section{Introduction}
	A figure of merit for hard ferromagnets is proportional to the magneto-crystalline anisotropy energy (MAE), which measures the energy cost of deviations from easy-axis magnetization. Typical strong magnets consist of rare-earth and transition-metal intermetallic compounds. The combination of strong spin-orbit coupling and large ordered moment gives rise to strong anisotropy. Due to the scarce availability of rare-earths, hard magnets without rare-earths are desirable. A candidate material consisting of 3$d$ and 5$d$ transition metals offers a platform for further exploration. \cite{mae-intro1,mae-intro2,mae-intro3} \\
	
	Transition metal dichalcogenides (\ce{MX_2}, with M = transition metal, X = S,Se,Te) form layered structures. They show interesting properties such as charge density waves and superconductivity \cite{dichal1,dichal2,dichal3,dichal4}. These properties can be tuned and enhanced by intercalating them with metal ions, resulting in changes in superconducting transition temperatures, and anisotropic magneto-transport \cite{intercalate1,intercalate2,intercalate3,intercalate4,
	intercalate5,intercalate6,intercalate7,Will}. 
\ce{Fe_xTaS_2} is an example where Fe ions are intercalated between the hexagonal layers of \ce{2H-TaS_2}. The magnetic properties can be varied with the concentration of Fe: it is found that for $x < 0.4$, the magnetic ordering is ferromagnetic, and switches to being antiferromagnetic for higher concentrations of Fe \cite{x0.4}. The Curie temperature, $T_C$ is highest for $x\!=\!0.25$ with a value of $160K$.\cite{Emilia} \\	

\ce{Fe_{0.25}TaS_2} has been observed to display large magneto-crystalline anisotropy, sharp switching of magnetization, and anomalous magnetoresistance \cite{Emilia}. Fe crystallizes in a $2 \times 2$ superlattice within the \ce{TaS_2} layers, with Fe ions forming a hexagonal crystal.\cite{superlattice1,main} . The arrangement is crucial as it allows RKKY interactions to maximize $T_C$ and also results in a large uniaxial magnetic anisotropy. For easy axis magnetization along the hexagonal $c$-axis, the hysteresis loop is almost square,\cite{Emilia} reminiscent of strong permanent magnets. On the other hand,  measurements in an in-plane magnetic field barely show discernible magnetization. An easy axis moment of $\sim 4\mu_B$ is found per Fe ion, of which an unusually large $1\mu_B$ was found to arise from the orbital component by previous X-ray magnetic circular dichoism (XMCD)  measurements\cite{main}. The large easy-axis orbital moment accounts for a calculated MAE of 15meV/Fe\cite{main}, a value comparable to rare-earth magnets.\cite{dft-mae4} \ce{Fe_{0.25}TaS_2} is thus a candidate hard magnet without rare-earth elements. \\

First-principles electronic structure calculations based on density functional theory (DFT) have long been used to estimate MAE values\cite{dft-mae1,dft-mae2,dft-mae3,dft-mae4,JianXin}. Typically, separate calculations are done with magnetic moment along and away from the preferred easy-axis direction. The energies can be compared by the force theorem\cite{force1} and total-energy difference methods \cite{dft-mae3}. 
 A less computationally intensive method was found with the so-called torque method, which involves restricting the magnetization to the $45^\circ$ angle to the easy axis and evaluating the angular derivative of the energy (torque)\cite{torque1,torque2}. Previous first-principles calculations have used the orbital-polarization scheme to account for many-body correlational effects.\cite{orb-pol1,orb-pol2,orb-pol3} \\ 

  It is essential to incorporate the many-body effects of electron correlations into the DFT calculations to reproduce the observed orbital moments, which are critical to the anisotropy energy. Here, we employ the DFT+U method \cite{dft+u1}, which provides a better description of the electron correlations. We study the magnetization and  anisotropy energy in \ce{Fe_{0.25}TaS_2}. We break down the dependence of MAE into its constituent elements, disentangling the effects of the crystal field environment and of spin-orbit coupling. In contrast to earlier first principles calculations\cite{main}, we find that a modest value of the Hubbard  on-site interaction $U\gtrsim 2.5$~eV is sufficient to reproduce the measured magnetic moment and the results of the X-ray absorption spectroscopy (XAS)\cite{main}. To further elucidate the origin of the large MAE in \ce{Fe_{0.25}TaS_2}, we study the substitutions of Fe and Ta by other $3d$ and $4d$ elements. Contrary to the intuitive expectation, we find that the presence of $5d$ electrons of Ta does not provide a significant source of MAE. Rather, it is the $d^6$ configuration of Fe$^{2+}$ ions that results in a large orbital moment and is thus responsible for the large observed MAE in \ce{Fe_{0.25}TaS_2}. Our findings bear important ramifications for the search for strong permanent magnets, in particular without rare-earth elements\cite{rare-earth}.
	
	
\section{Methods}	
	We performed first-principles calculations within the DFT+U scheme in the generalized gradient approximation (GGA-PBE) for exchange-correlation \cite{pbe}. The full-potential, linearized augmented plane wave (FP-LAPW) method was used as implemented in the Wien2K code\cite{wien2k}. A $13 \times 13 \times 6 $ Monkhorst-Pack $\mathbf{k}$-point grid was used for BZ integration with the tetrahedron method. We performed a range of calculations by varying the Hubbard on-site energy parameter, $U$ on the Fe site. The Hund exchange parameter, J was fixed to 0.7eV.
	
	To choose the optimal value for Hubbard $U$, the calculated moments were compared with the experimental values. To examine the anisotropy, calculations were done by restricting the magnetization along the easy axis (001) and along a hard direction in the basal plane (010). Noting that spin-orbit coupling(SOC) is the main contributor, MAE was  calculated in two ways: 
	(a) as a difference in values of SOC energy between the two directions: $E_{MAE} = \Delta \langle \zeta L\cdot S \rangle $ ($\zeta$ being the SOC constant); and 
	(b) an approximation involving the orbital moment anisotropy as: 
	$E_{MAE} = \tfrac{1}{4} \zeta \langle\Delta L\rangle \cdot \langle S\rangle $, 
	where $\langle\Delta L\rangle$ is the change in orbital moment due to crystalline anisotropy \cite{mae_formula}. We found both methods to give consistent results and to be much more accurate than the brute-force comparison of total energies of the two spin configurations. The error in the total energy difference is an artifact of the Hubbard interaction term in DFT+U. This term is linear in $U$, and involves pair-wise products of orbital occupations.\cite{dft+u1} As will be explained later, the orbital occupations are significantly different between the two magnetic directions. This gives rise to a misleading U-dependent correction when MAE is calculated as the total energy difference, one that is an order of magnitude larger than the SOC effect. 
	
	
\section{Results}	
\subsection{Selection of U}
	
\begin{figure}[h]
	\includegraphics[scale=1]{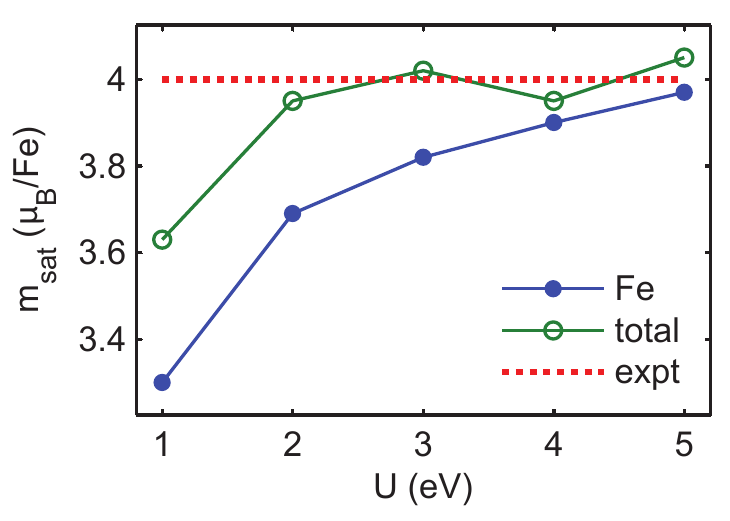}
	\caption{ \label{fig_moment_Fe_tot}
	Comparison of saturated moment from entire cell (green, empty circles) with experimental value shows $U \gtrsim 2.5eV$ to be sufficient. Restricting to only moments within the Fe sphere (blue, filled circles) would require larger U to match experiment.
	}
	
\end{figure}

	Bare DFT calculations (without Hubbard U interaction) yielded a ferromagnetic ordering with spin moment, $m_s=3.2\mu_B$/Fe and orbital moment, $m_o=0.3\mu_B$/Fe. These values can be compared with the moments concluded from XAS measurements\cite{main}. While $m_s$ agrees with the experimental value of $3\mu_B$, $m_o$ is largely underestimated from the experiment value of $1\mu_B$. In order to cure this deficiency of the DFT, we have performed a series of  DFT+U calculations. As the Hubbard interaction $U$ on Fe site was progressively increased, $m_o$ rose to $0.7\mu_B$ and plateaued beyond $U\sim 2.5eV$.~\ref{fig_MAE_U} We plot the saturated moment ($m_{\text{sat}}=m_s+m_o$) in Fig.~\ref{fig_moment_Fe_tot} (green, empty circles). On comparing the calculated saturated moment of the system with the experimental value($m_{\text{sat}}=4\mu_B$)\cite{Emilia}, we find $U\sim 2.5$~eV to be sufficient. \\
	
	We note that the previous first-principles study\cite{main} was performed using a higher value of $U = 4.5$~eV, which is rather larger than the typically expected value in other correlated Fe compounds.\cite{fe-dftu1,fe-dftu2} We find that the justification by the authors of Ref.\onlinecite{main} for using such a large value of $U = 4.5$~eV was likely a result of attempting to match the experimentally observed magnetic moment with the calculated moment exclusively on the Fe-site (blue, filled circles in Fig.~\ref{fig_moment_Fe_tot}). However, it is often the case that the interstitial regions outside of the muffin-tin sphere provide a non-negligible contribution to the magnetic moment, which is indeed the case here (green, empty circles in Fig.~\ref{fig_moment_Fe_tot}). We conclude that it is therefore important to the include the magnetic moments in interstitials into the calculation. We note that this contribution is largest to the spin moment (interstitials contribute $\sim 7\%$ of the total $m_s$ value), and is negligible for the orbital moments (less than $1.5\%$ of $m_o$). We conclude that the Hubbard $U\gtrsim 2.5$ eV is sufficient to reproduce the experimentally measured saturated moment\cite{Emilia}. 
	
\begin{figure}[h!]
	\includegraphics[scale=1]{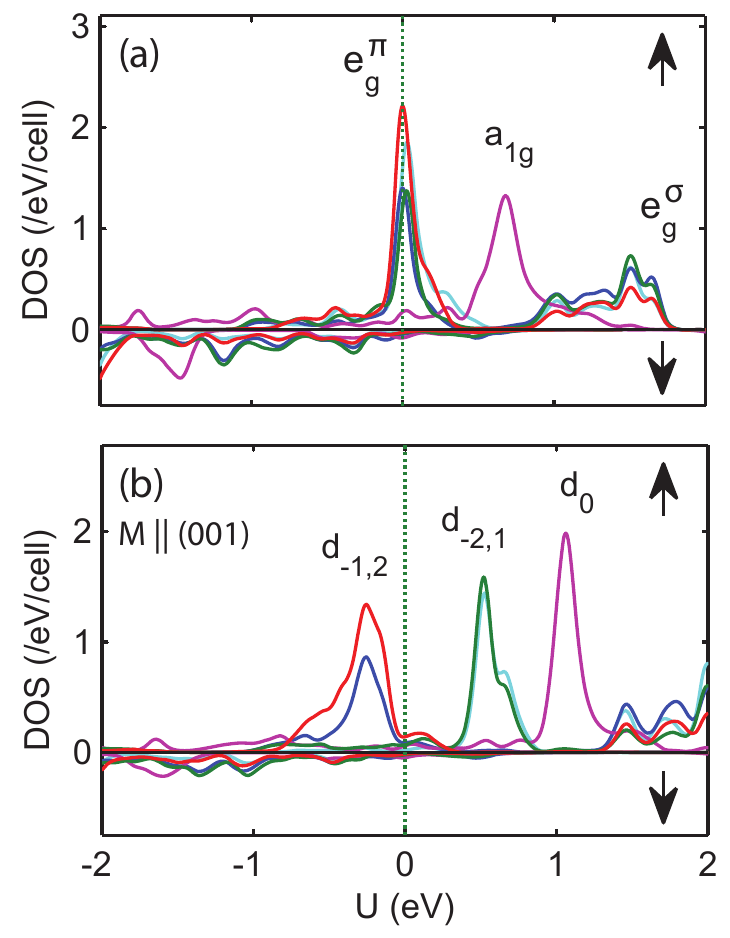}
	\caption{\label{fig_DOS_U+SOC}
	DOS projected onto Fe d-orbitals. (a) Insufficient Hubbard interaction U ($U \le 2eV$), and lack of SOC; (b) Sufficient U (2.5 eV) with SOC (moment quantization along the easy axis - (001)). Majority spin (bottom panels) and minority spin (top panels) contributions are shown. The colors represent the different $d_m$ orbitals, the relevant orbitals are labeled. \\
	 Inclusion of Hubbard U and spin-orbit coupling(SOC) is necessary to reproduce experimental occupation}
\end{figure}

A further check on the value of Hubbard interaction $U$ can be performed by comparing the calculated easy-axis orbital occupations and splittings with the XAS measurements \cite{main}. While the majority spin orbitals at the Fe site are fully occupied, XAS measurements show that the minority spin is occupied mainly by orbitals $d_m$ with orbital momentum projection $m=-1$ and $m=2$.
This anisotropic distribution accounts for the large orbital moment on Fe site. To compare with the XAS results, we plot in Fig. \ref{fig_DOS_U+SOC} the electron density of states (DOS) projected onto Fe $d_m$-orbitals. Figure \ref{fig_DOS_U+SOC}a) shows the density of states calculated in the absence of SOC and for small Hubbard $U$. The orbital splitting corresponds to the $D_{3d}$ trigonal symmetry, caused by the distorted octahedron coordination of S atoms around the Fe site. The symmetry being lower than the cubic case leads the Fe 3d-orbitals to split into $e_g^\pi$, $a_{1g}$ and $e_g^\sigma$ sub-bands.
In this case, the $e_g^\pi$ orbitals (consisting of $m=\pm1, \pm2$) dominate the DOS at the Fermi level. This however does not match the XAS results which find the $e_g^\pi$ multiplet split. \\

It turns out that in order to reproduce the XAS findings, both the spin-orbit coupling and a sufficiently large Hubbard U ($U\ge 2.5eV$) must be included in the calculations. In this case, we find that the $e_g^\pi$ states at the Fermi level split into lower-energy $m=-1,2$ states (as in XAS experiment\cite{main}) and higher-energy unoccupied $m=-2,1$ states, as illustrated in Fig.~\ref{fig_DOS_U+SOC}b.
As the value of Hubbard $U$ is increased beyond 2.5eV, these $d_{-1}/d_2$ orbitals lower further. If the energy separation between the occupied and unoccupied parts of the $e_g^\pi$ orbitals were known experimentally, it would  would help ascertain the value of Hubbard U more precisely. For the purpose of this work, in what follows we shall focus on $U = 2.5eV$, which also reproduces the experimentally measured saturated magnetic moment in Fig.~\ref{fig_moment_Fe_tot}, as already mentioned above. \\
	

\subsection{Anisotropy energy}	

\begin{figure}[h]
	\includegraphics[scale=1]{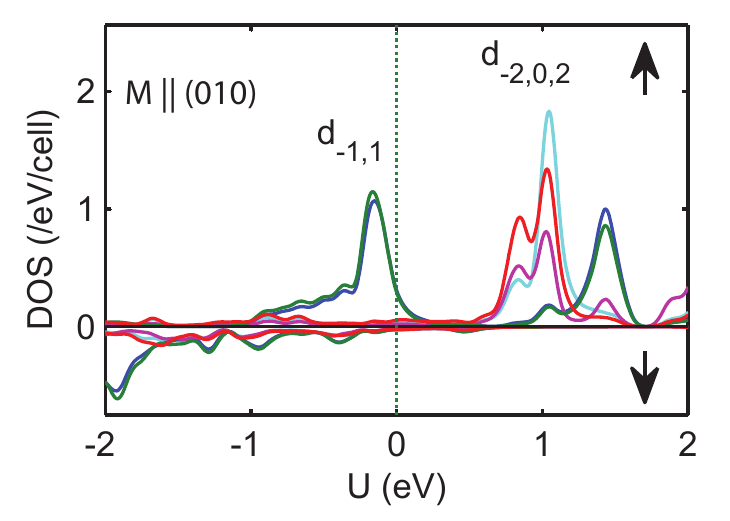}
	\caption{DOS projected onto Fe d-orbitals for hard-axis magnetization(010). Majority spin (bottom panels) and minority spin (top panels) contributions are shown. The colors represent the different $d_m$ orbitals, the relevant orbitals are labeled. \\ Occupations differ from the easy-axis case, leading to orbital anisotropy.}
	\label{fig_DOS_basal_plane}
\end{figure}

\begin{figure}[h]
	\includegraphics[scale=1]{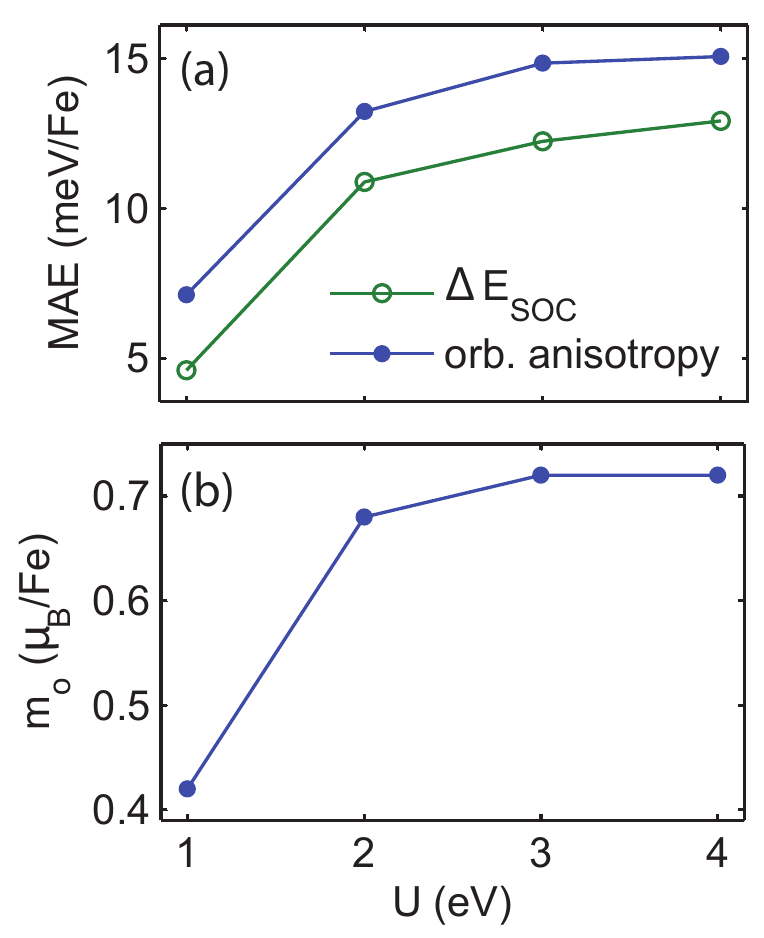}
	\caption{(a) MAE calculations from $\Delta E_{SOC}$ (green, empty circles), and orbital anisotropy (blue, filled circles). (b) Easy-axis moment shows an analogous trend }
	\label{fig_MAE_U}	
\end{figure}

In our DFT+U calculations with $U=2.5$~eV, we obtain the spin and orbital moment $m_s = 3.3\mu_B$ and $m_o = 0.7\mu_B$, respectively, resulting in the total moment on Fe site $m_{tot} = 4\mu_B$. Although the total moment matches the experimental value\cite{Emilia}, our value of $m_o$ is less than the previous calculated and measured value\cite{main} of $1\mu_B$. This difference could be attributed to the technical differences in the implementation of the DFT+U method.\cite{dftu-dcc}
The occupation analysis of $d$-levels shows that Fe ions have a +2 oxidation state, resulting in a high-spin $d^6$ configuration. We next performed similar calculations with the moment along the magnetic hard direction in the basal plane. Such moment arrangement lowers the hexagonal symmetry, resulting in a different splitting of Fe $e_g^\pi$-orbitals, with $d_{\pm1}$ becoming lower in energy (fig \ref{fig_DOS_basal_plane}). This in turn leads to a lower orbital moment, $m_o = 0.1\mu_B$, while the spin moment is same as earlier. 
The resulting large anisotropy in the magnitude of the orbital moment, $\Delta L = 0.6 \mu_B$, is crucial for the large magnetocrystalline anisotropy. MAE can be estimated from the orbital anisotropy as 
$E_{MAE} = \tfrac{1}{4} \zeta \langle\Delta L\rangle \cdot \langle S\rangle $
 (here $\zeta$ is the spin-orbit coupling constant). The resulting MAE dependence on the strength of Hubbard repulsion $U$ is plotted in Fig.~\ref{fig_MAE_U}(a) (blue, filled circles).

To verify the validity of these results, we have also computed MAE from the  variation of the SOC energy, $\Delta E_{SOC}$ between the easy-axis and hard magnetic direction\cite{mae_formula} (filled circles in Fig.~\ref{fig_MAE_U}). Both sets of data are in reasonable agreement with one another (within $15\%$) and both show the saturation of MAE as $U$ exceeds $\sim 3$~eV. The Hubbard interaction, along with SOC split the orbitals as explained in the previous section. The resulting variation in orbital moment, $\Delta L$ is responsible for the anisotropy. The two panels in Fig.~\ref{fig_MAE_U} show the correspondence between MAE and the (easy-axis) orbital moment, $m_o$. We estimate the value of MAE at $U=2.5$~eV as the average of the two methods, yielding $E_{MAE}\approx 12$~meV per Fe. This large value of MAE is comparable to those found in rare earth magnets\cite{dft-mae4}. The corresponding anisotropy field, required to rotate the magnetization from the easy to hard direction, can be found from $E_{MAE} = \mu\cdot B_{ani}$ (where $\mu = \mu_B(L + 2S)$ is the saturated magnetic moment), yielding $B_{ani}\approx 62$~T, consistent with the experimental estimate of 60~T \cite{main}. 
	 
\subsection{Origin of MAE}
Naively, one would expect the 3d-electon metal to provide magnetic moment, and its interaction with a 5d-electron metal to provide anisotropy.
We put this hypothesis to test by studying the dependence of the MAE on the constituent elements, Fe and Ta.
Below, we report on two different approaches: we first study the substitution of the individual elements in the material and its effect on MAE. Next, we artificially tune the strength of spin-orbit coupling on Fe and Ta sites independently of one another, in order to disentangle the relative contribution of these sites to the MAE.
	

\subsubsection{Elemental substitution}

The importance of each of the constituent elements can be examined by substituting them with another carefully selected element. To test the strong SOC coupling effect due to Ta, it can be replaced by one with lower SOC strength. Nb, the element immediately above Ta in the periodic table preserves the crystal structure and the electronic configuration in \ce{Fe_{0.25}NbS_2}. Nb, being a $4d^3$ element is expected to have lower SOC strength than the $5d^3$ element, Ta. Within the DFT+U implementation, we calculate the SOC constant for Nb($\zeta_{Nb} = 107$meV) to be smaller than Ta by a factor of four($\zeta_{Ta} = 417$meV). Next, to test the effect of the electronic configuration of Fe, we replaced it with the neighboring element, Mn to study \ce{Mn_{0.25}TaS_2}. The minority spin electron in $3d^6$ \ce{Fe^{2+}} ion, which was responsible for orbital moment is absent in $3d^5$ \ce{Mn^{2+}}.
(We also attempted to substitute Fe with other 3d elements - Cr, Co and Ni. However, unlike Mn, these elements did not preserve ferromagnetic ordering.) The MAE calculations were performed on \ce{Fe_{0.25}NbS_2} and \ce{Mn_{0.25}TaS_2}.

\begin{table}[h]
	\centering
	\begin{tabular}{|c|c|c|c|c|} \hline
		Material & \multicolumn{2}{|c|}{MAE splitting (meV)} &  $m_o(\mu_B)$ &  $m_s(\mu_B)$ \\ \hline
		\ce{Fe_{0.25}TaS_2}	& Fe : 14 & Ta : -1.0  & 0.7 & 3.3 \\ \hline
		\ce{Fe_{0.25}NbS_2}	& Fe : 12 & Nb : -0.1 & 0.6 & 3.2 \\ \hline
		\ce{Mn_{0.25}TaS_2}	& Mn : 0 & Ta : -0.1 & 0.01 & 4.1 \\ \hline
	\end{tabular}
	\caption{Comparison of MAE and the orbital ($m_0$) and spin moments ($m_s$) on the $3d$ element site for different compounds. (Note that there are 4 times as many Ta/Nb atoms as Fe/Mn, so one would expect a larger effect on MAE by substituting on Ta site.)}
	\label{table_substitution}
\end{table}

Table \ref{table_substitution} shows the contribution to MAE from the constituent elements. In \ce{Fe_{0.25}TaS_2}, a major part of the MAE value arises from Fe. The four Ta atoms together account for a small negative value. (This means the easy and hard directions are switched for Ta moments.) On substitution of Ta with Nb, the MAE and $m_o$ on Fe decrease slightly, while Nb accounts for an even smaller contribution to MAE than Ta. A four factor reduction in SOC strength from Ta to Nb, only contributes a $15\%$ decrease in MAE. On the other hand, substitution of Fe with Mn results in negligible MAE and orbital moment in \ce{Mn_{0.25}TaS_2}. We conclude that the $d^6$ electron configuration of Fe was crucial for the resultant orbital moment and MAE. These results show that MAE mainly arises from the Fe site. A partial contribution to MAE is caused by the influence of Ta on the Fe moments. 
	

\subsubsection{Scaling of the SOC constants}

\begin{figure}[]
	\includegraphics[scale=1]{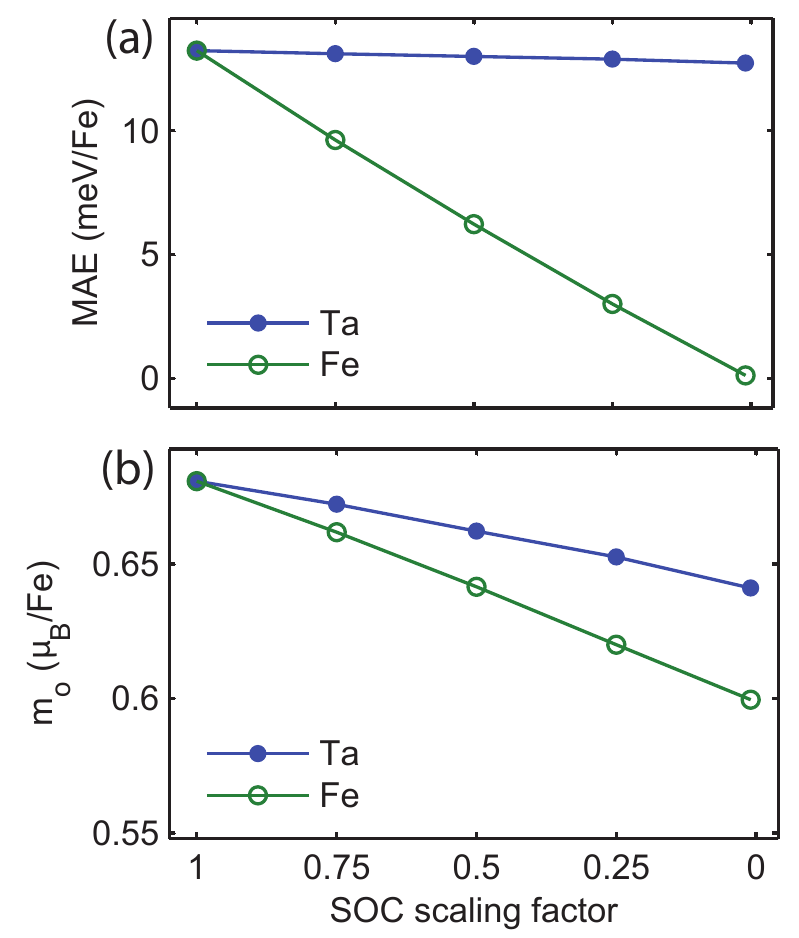}
	\caption{Dependence of (a) orbital moment, and (b) MAE on the SOC strengths on different atoms. Blue, filled (green, empty) circles indicate varying $\zeta$ on Ta (Fe) only. 
	}
	\label{fig_SOC_scaling}	

\end{figure}

The breakdown of MAE can further be explored by artificially scaling down the SOC strengths on Fe and Ta atoms. SOC gives rise to the orbital moment, and is responsible for MAE. We were able to scale down the SOC strengths on Fe and Ta independently. This was done by artificially decreasing the SOC constant, $\zeta$ at each site. The effect was examined by inspecting changes to MAE and $m_o$. Fig \ref{fig_SOC_scaling} (a) shows the changes to MAE as the SOC constant is gradually reduced on each of the atoms selectively. MAE is seen to vanish as the SOC strength on Fe is diminished (green, empty circles), similar to Mn-substitution from the previous section. On the other hand, the change in MAE due to the SOC strength on Ta is an order of magnitude smaller (blue, filled circles), comparable to Nb-substitution from the previous section. \\

We also plot the orbital moment, $m_o$ variations in Fig \ref{fig_SOC_scaling} (b). $m_o$ decreases linearly on reducing the SOC strength at either of the sites. The orbital moment is affected twice as much due to SOC strength at the Fe site (green, empty circles), compared with Ta site (blue, filled circles). As MAE depends on both orbital moment and SOC strength, MAE drops more rapidly. This set of calculations further confirm what we found in the previous section. 
As was shown earlier, the combination of Hubbard interaction and spin-orbit coupling results in the orbital anisotropy. In this section, we fix Hubbard $U$ to a sufficiently large value ($U\sim 2.5$ eV), and only alter the SOC strength. Starting from a state with large orbital moment, like in Fig.\ref{fig_DOS_U+SOC} (b), the decreasing SOC strength measures deviations to orbital moment. In the limit of vanishing SOC strength, $m_o$ is non-zero as the orbital splitting in Fig.\ref{fig_DOS_U+SOC} (b) is preserved. The same test performed with a small $U(=1$ eV) shows $m_o$ to vanish as SOC strength is decreased to zero(not shown here). In this case, the lack of splitting of states at the Fermi level as seen in Fig.\ref{fig_DOS_U+SOC} (a), produces a state with lower orbital moment. As SOC strength is decreased, orbital moment gets quenched.\\
	 
The moments in \ce{Fe_{0.25}TaS_2} lie mainly at the Fe site. The crystal field splitting and coupling with the lattice result in a highly anisotropic orbital moment on Fe. This is the main contributor to total MAE. Although the 5d electrons in Ta have a large SOC strength, the lack of magnetic moment at the Ta site results in a much smaller contribution to the total MAE. Their interactions with the Fe electrons only partly enhances the MAE.
	

\section{Conclusion}
\ce{Fe_{0.25}TaS_2} is an interesting magnetic material composed entirely of transition metals. It shows magnetic properties comparable to rare-earth based magnets. We have investigated its magnetocrystalline anisotropy from DFT+U based first-principles calculations. We observe a large uniaxial MAE that arises due to the orbital anisotropy at the Fe site. The Coulomb interaction strength and crystal field splitting of the Fe ion give rise to the orbital moment anisotropy. We were able to test the effect of spin-orbit coupling strength of the heavier element in the dichalcogenide layer on the material's MAE. Subsituting Ta with a lighter element such as Nb greatly reduces the spin-orbit strength. We also artificially vary the spin-orbit constant of the heavier element, and note the changes in orbital moment. In both cases, we find the changes to MAE and orbital moment to be smaller than previously expected. The dichalcogenide layer lacks magnetic moment, and has no direct overlap with the magnetic intercalant ion. So, its spin-orbit strength weakly affects the anisotropy produced by the intercalant. Instead, the dichalcogenide layer plays an important role in forming a favorable crystal field environment for the intercalant ion. Our analysis helps in the search of strong permanent magnets with transition metals. It would be interesting to study systems where the dichalcogenide layer consists of lighter $3d$-electron elements. \\
	

\section*{Acknowledgements}
One of the authors (V.L.) acknowledges the hospitality of Los Alamos National Laboratory (LANL), where part of this work was carried out. This work  was supported by Welch Foundation grant C-1818 (V.L. \& A.H.N.), and the U.S. DOE at LANL under Contract No. DEAC52-06NA25396 through the LANL LDRD Program (V..L. \& J.X.Z.).	
	

\bibliographystyle{phaip} 

\begin{thebibliography}{99}
\bibitem{dftu-dcc} Various implementations of the DFT+U method differ in the way they account for the double-counting correction. While we used the Self-interaction correction form of DFT+U, we speculate the previous authors may have used the Around Mean-field form. Our checks with the second method resulted in a higher $m_o$ value of $0.8\mu_B$.
\end{thebibliography}

\end{document}